\documentclass{iaus}
\usepackage{graphicx}

\def\tex {\ifmmode{{T}_{\rm ex}}\else{$T_{\rm ex}$}\fi}
\def\tmb {\ifmmode{{T}_{\rm mb}}\else{$T_{\rm mb}$}\fi}
\def\ci     {\ifmmode{{\rm C}{\rm \small I}}\else{C\ts {\scriptsize I}}\fi}
\def\hi     {\ifmmode{{\rm H}{\rm \small I}}\else{H\ts {\scriptsize I}}\fi}
\def\hh     {\ifmmode{{\rm H}_2}\else{H$_2$}\fi}
\def\ts     {\thinspace}
\def\kms    {\ifmmode{{\rm \ts km\ts s}^{-1}}\else{\ts km\ts s$^{-1}$}\fi}
\def\msol   {\ifmmode{{\rm M}_{\odot}}\else{M$_{\odot}$}\fi}
\def\lsol   {\ifmmode{{\rm L}_{\odot}}\else{L$_{\odot}$}\fi}
\def\zsol   {\ifmmode{{\rm Z}_{\odot}}\else{Z$_{\odot}$}\fi}
\def\etal   {{\rm et\ts al.}}

\title[Molecular gas] %% give here short title %%
{Molecular gas in galaxies at all redshifts}

\author[F. Combes]   %% give here short author list %%
{Francoise Combes}

\affiliation{Observatoire de Paris, LERMA and CNRS,
61 Av. de l'Observatoire, F-75014 Paris, France \\ email: {\tt francoise.combes@obspm.fr} }

\pubyear{2011}
\volume{277}  %% insert here IAU Symposium No.
\pagerange{111--118}
\jname{Tracing the ancestry of galaxies (on the land of our ancestors)}
\editors{C. Carignan, F. Combes, K. Freeman, eds.}
\begin{document}

\maketitle

\begin{abstract}
I review some recent results about the molecular content
of galaxies, obtained essentially from the CO lines, but also
dense tracers, or the dust continuum emission. New results have been
obtained on molecular cloud physics, and their efficiency
to form stars, shedding light on the Kennicutt-Schmidt law
as a function of surface density and galaxy type.
 Large progress has been made on galaxy at moderate and high redshifts,
 allowing to interprete the star formation history and star formation
efficiency as a function of gas content, or galaxy evolution.
  In massive galaxies, the gas fraction was higher in the past, and
galaxy disks were more unstable and more turbulent.
 ALMA observations will allow the study of
more normal galaxies at high z with higher spatial resolution and sensitivity.
%% add here a maximum of 10 keywords, to be taken form the file $<$Keywords.txt$>$
\end{abstract}

\firstsection % if your document starts with a section,
              % remove some space above using this command.

\section{Nearby galaxies}

A new survey (HERACLES) has been completed with the IRAM receiver array 
in the CO(2-1) line,
allowing extended maps of nearby galaxies, at 12'' resolution
(Leroy et al 2009). The survey contains an atlas of 18 nearby galaxies,
observed at multi-wavelengths, and in particular in the HI line, and in the 
mid infrared by Spitzer. 
Among the results, it is interesting to note a
very good correlation between CO and HI kinematics. The excitation of
the molecular gas, as traced by the first two rotational lines of CO, 
is usually low in the disk (R = CO(2-1)/CO(1-0) = 0.6) while it is higher in nuclei
(R=1), indicating denser gas. The CO emission is 
compatible with optically thick clouds at a kinetic temperature of T=10K.

A more refined view of the star formation law in galaxies has been
obtained by Bigiel et al (2008). The Schmidt-Kennicutt law relating
star formation and gas density has a different slope $n$, according to the gas
surface density. At high surface density, when the gas is molecular,
the gas forms stars at a constant efficiency ($n$=1), and the 
time-scale for  star formation is 2 10$^9$ yrs. While, at low
surface density, when the gas is atomic, the slope is much higher
$n$= 2 or more. At sub-kpc scale, the star formation rate is
not strongly correlated with HI surface density. The transition between
HI and H$_2$ occurs when the surface density is $>$ 9 M$_\odot$pc$^{-2}$.

In order to better determine the change across the spiral arms
of the molecular gas physical properties, Schinnerer et al (2010) have
mapped at interferometric resolution several lines of CO and its isotopes,
together with dense gas tracers, such as HCN and HCO$^+$, in two selected regions
across M51 spiral arms. They find no change across the arms, and 
the GMC population in the spiral arms of M51 is similar to those of the Milky Way,
even if the star formation rate is much higher. 

On the contrary, the low surface brightness dwarf spiral
M33 reveals different conditions than in the Milky Way. In the center,
the lower metallicity is the cause of a higher conversion factor, as will be described
by J.Braine in this conference (see Gratier et al 2010), while surprisingly
the outer molecular complexes show exceptionally bright CO  emission,
and therefore a lower conversion factor (Bigiel et al 2010).

The excitation of the molecular gas has also been investigated
with the first 3 CO rotational lines, at high resolution
with the PdB and SMA interferometers in two NUGA-sample
galaxies (Boone et al 2010).
At about 100pc-scale resolution, it is possible to distinguish warmer
and less dense gas towards the center, may-be heated by the AGN,
and colder and denser components in a circumnuclear arc.  According
to LTE analysis, more then 50\% of the gas is optically thick in both 
galaxies.

\begin{figure}[b]
% \vspace*{-2.0 cm}
\begin{center}
 \includegraphics[width=9.2cm]{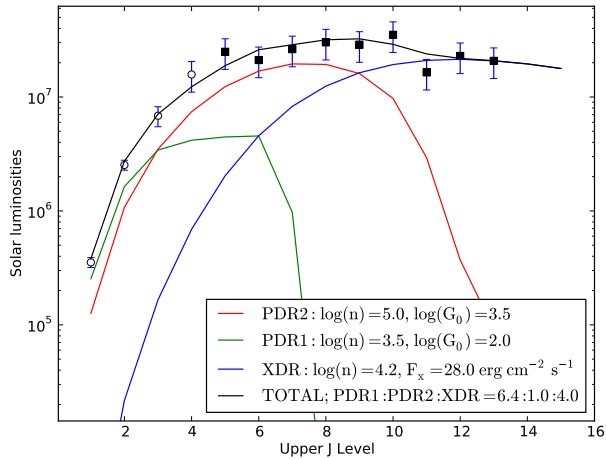} 
% \vspace*{-1.0 cm}
 \caption{
Energy distribution in the various CO lines from Mrk231:
the high frequency measurements (filled symbols) have been done with SPIRE
on Herschel, while the low frequency ones (open symbols) are measured from the
ground. The data can be reproduced by the combination of three models: two
model PDR components  (red and green lines) and an XDR component (blue line). 
The sum of these three
  components is the black line, made to fit the CO measurements. The PDR alone
are not sufficient.
From van der Werf et al (2010).}
   \label{fig1}
\end{center}
\end{figure}

The CO excitation has recently ben estimated even more completely
by determining the spectral line energy distribution or SLED in
bright starbursts and quasars, with the Herschel Spire FTS instrument.
It is possible to obtain the full spectrum at once, although with
low spectral and spatial resolution, up to the CO(13-12) line.
In Messier 82, the SLED obtained by Panuzzo et al (2010)
reveals a significant part of the molecular gas at high temperature T=500K,
where the H$_2$ lines are the main coolant. The peak of the CO line intensity
occurs at the level $J$=7. 
Extremely high CO line excitation is a clue to the presence of an AGN and 
its strong X-ray heating (XDR). The star formation region are characterized
by PDR, where dust is heated efficiently. Also there is a
richer chemistry in XDR (H$_2$O, H$_2$O+, OH+..).
Figure \ref{fig1} reveals the CO SLED of Mrk231, a typical
quasar, where the excitation is dominated by an XDR.

For more common lower-energy AGN, the XDR is visible only
very close to the nucleus.
The ALMA spatial resolution is then required to resolve these regions.
A typical example is the Seyfert 2 galaxy NGC 1068,
where the XDR dominates the starburst regions only
at r $<$ 70pc.
Garcia-Burillo et al (2010) have mapped 
SiO, CN with the IRAM interferometer, which are
tracers of shocks as well as CH$_3$OH, HNCO.

The starbursts in barred galaxies are frequently found
in nuclear rings, at Lindblad resonances, such as in NGC1097.
The recent observation with Herschel of the dust morphology in
this ring by Sandstrom et al (2010) has revealed a high
uniformisation of the dust heating in the ring, suggesting 
some kind of smoothing, or that the
ISRF is a significant source of heating.
This smooth structure of the dust contrasts with the clumpy nature
of the line emission, in the infrared lines: [OI] 63$\mu$,
[OIII] 88 $\mu$, [NII] 122 $\mu$, [CII] 158 $\mu$  and [NII] 205 $\mu$
(Beirao et al 2010).

Galaxy formation models required strong feedback to limit star formation.
This feedback is observed as molecular gas outflow out of starbursting
region, such as the prototypical dwarf galaxy M82. Recently, gas outflows
have been observed also generated by AGN feedback, in the nearby quasar
Mrk 231, with OH lines and outflow velocities up to 1400 km/s
(Fischer et al 2010).
In the same quasar, the observation of the CO(1-0) line with the IRAM
interferometer revealed broad wings, with velocities up to 750km/s, dragging
the molecular gas in strong outflows (Feruglio et al 2010). According to
this interpretation, most of the gas around the nucleus could be depleted in
a time-scale of tens of Myrs.
In Mrk 231, Gonzalez-Alfonso et al (2010) have reported about emission 
and absorption lines of H$_2$O, which is abundant due to shocks, or XDR
chemistry, and evaporation of ice from grains.
A clear example of outflows has also been seen in 4C 31.04 (Garcia-Burillo et al 2007).
 The HCO$^+$ profile is very wide, broader than 1000km/s. There is both emission and
deep absorption in the blue-side, and an interferometer map has been able
to map both, according to its spatial coincidence with the resolved jet in radio
continuum. 
Another striking example, is the radio source
 3C293, observed in HI absorption with a width of 1400km/s (Morganti et al 2003).
The absorption is blue-shifted, indicating an outflow.
Garcia-Burillo et al (2010) have mapped the source in CO and HCO$^+$, where 
a deep absorption is also detected.

The strongest evidence of AGN feedback until now has been seen 
in the center of cool core clusters. Cold molecular gas has been detected
in the CO lines, associated to the cooling flow in Perseus (Salome et al 2006).
Recently, the cooling lines OI and CII have been mapped with 
PACS and SPIRE in some cooling flows (Edge et al 2010, Mittal et al, in prep).
They were seen with the same morphology, and when possible 
with the same kinematics than for the CO lines. They appear to 
come from the same gas, cooling through different phases,
showing no rotation, but in(-out)flows.

A strong result from the first Science with Herschel is
the evidence of a cold dust component in dwarfs
and the outer parts of spiral galaxies (Grossi et al 2010).
 This has been confirmed with Planck, and also
ground based bolometer, such as LABOCA.
In dwarfs, the CII/CO ratio is very high, due to low-metallicity
effects (Cormier et al 2010): the high UV environment,
due to the lack of dust, provides large-scale photodissociation of
the molecular gas. However, the SED of dust emission reveals and
overabundance of cold dust,  or a flat opacity law $\beta <$1.5 in low-Z systems
(Boselli et al 2010). This is the case in the LMC (Meixner et al 2010),
and also M33 (Quintana-Lacaci et al. in prep).
A 2-component grey body fit with $\beta$=2 indicates a very cold
component at 5.7K, 15 times more massive than the 21K
component. This excess is clearly seen in the low-Z dwarf
NGC 1705 (O'Halloran et al 2010).

\section{Galaxies at high redshift}

In the last decade, a large number of CO detections have
occured at high redshift, and surprising is the very
good correlation between the FIR and CO fluxes, even 
for quasars where the AGN contribution to the FIR 
might be significant (Iono et al 2009).
In the early universe, about a dozain of objects at z$>$ 4 have been detected in CO,
but they are all lens  amplified by a large factor (Wang et al 2010, Riechers et al 2010a)
The most distant object is the QSO at z=6.4 (Fan et al 2003).
The mass of dust is 10$^8$M$_\odot$ (Bertoldi et al 2003), and its black hole 
mass 1.5 10$^9$M$_\odot$ (Willot et al 2003). The CII
line has been detected by Walter et al (2009), and the data are consistent with a
1kpc-scale starburst, of SF density 1000M$_\odot$/yr/kpc$^2$.
Surprisingly HCN has not been detected in this source, although
HCN appears better correlated to star formation than CO (Gao \&
Solomon 2004).

HCN as a high density tracer has been detected at high z
tracing nH$_2$ $\sim$ 10$^5$ cm$^{-3}$. One of the strong detections
is the Cloverleaf  quasar at z=2.56, dominated by a starburst.
CO, HCN, HCO$^+$ trace the warm and dense gas forming stars
(Riechers et al 2010b).
In APM08279+5255, at z=3.91, the excitation is radiative in addition to collisional,
under the influence of the massive black hole (Riechers et al 2010c).
In some of the compact star forming regions,
the SFR is so high that the gas surface density might
be limited by the radiation. In other words, we are seing
Eddington limited star formation, at about SFR = 500-1000 M$_\odot$/yr/kpc$^2$.
The column density is in average NH$_2 \sim$ 10$^{24}$cm$^{-2}$,
100 times larger than in Giant Molecular Clouds (Riechers et al 2009a).

It is very difficult to detect the molecular component
of more ``normal'' objects, with  a moderate star formation rate.
This is only done with the help of gravitational lenses: the
``cosmic eye'' corresponds to a Lyman-break galaxy (LBG)
 at z=3.07 detected in CO(3-2) with the IRAM interferometer,
with a magnification factor of 28. Its SFR is 60 M$_\odot$/yr,
and its molecular content MH$_2$ = 2.4 10$^9$ M$_\odot$,
implying a starburst time-scale of 40Myr
(Coppin et al 2007).

\begin{figure}[b]
% \vspace*{-2.0 cm}
\begin{center}
 \includegraphics[width=8.2cm]{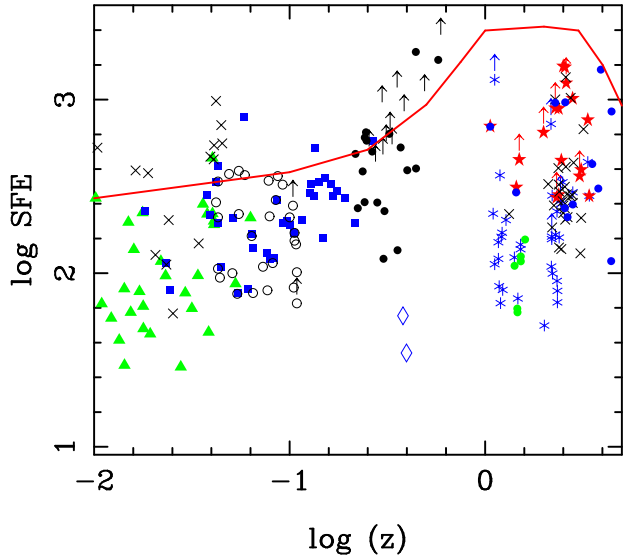} 
% \vspace*{-1.0 cm}
 \caption{
Evolution with z of the Star Formation Efficiency (SFE) defined by the ratio between
far-infrared luminosity and molecular gas mass (L$_{\rm FIR}$/M(H$_2$)),
for the sources detected between 0.2 $<$ z $<$ 0.6 (black full dots, and upward-arrows),
compared with other detections in the literature (cf Combes et al 2010). The red curve
is a schematic line summarizing the cosmic star formation history,
from the compilation by Hopkins \& Beacom (2006), implemented with the
GRB data by Kistler et al (2009), and the optical data from Bouwens et al (2008).}
   \label{fig2}
\end{center}
\end{figure}

Another category of sources frequently detected in molecules
are the SMG (Submillimeter Galaxies), discovered through
their dust continuum in blind surveys with bolometers.
 From this dust continuum, redshifted in the millimeter range,
it is possible to deduce their rest-FIR flux, which is an indicator
of star formation. The molecular content is derived
from the CO luminosity, after applying a conversion factor
adapted for ultraluminous galaxies (ULIRG). Their star formation efficiency,
as traced by the ratio between the FIR and CO luminosities is
strongly increasing with redshift (Greve et al 2005).
In average, their star formation rate is 700 M$_\odot$/yr,
with a starburst time-scale of 40-200 Myr.  SMG
appear more efficient to form stars than ULIRG. This
could be due to galaxy mergers, with objects not yet 
stabilized by bulges, the latter accumulating mass later.

Through selection effects, and because of the negative K-correction
in the dust continuum emission at high redshift, objects at $z>1$
have been much more studied than those at intermediate redshift.
To fill the gap, we have undertaken a search for CO emission
at $0.2 < z < 1$, selecting the brightest ULIRGs at these
redshifts (Combes et al 2010). One of the first results between $0.2 < z < 0.6$
is the increase of the gas content with z.
The detection rate is about 60\%, and the 
average H$_2$ mass is 1.6 $10^{10}$ M$_\odot$, 
taking the low CO-to-H$_2$ conversion factor assumed for ULIRGs. The maximum amount 
of gas available for a single galaxy is quickly increasing as a function of redshift. 
The star formation efficiency (SFE), traced by the ratio of FIR to CO luminosities, 
is found to be significantly higher, by a factor 3, than for local 
ULIRGs, and are comparable to high redshift ones  (Figure \ref{fig2}).
 The increase of SFE with z follows the increase of star formation.

ULIRG are often detected in the pure rotational lines of H$_2$ (e.g. by Spitzer).
 However, recently Zakamska (2010) has shown that most of the 
H$_2$ emission comes from outside the starburst region. Indeed, the H$_2$ lines do not
suffer dust obscuration, like the other tracers of SF, or PAH.
H$_2$ emission could be due to schocks outside SF regions, triggered
 by galaxy interactions. H$_2$ then
accelerates the cooling and collapse of the gas.

ULIRG are not all of high SF efficiency, there exists a category
of luminous galaxies with low efficiency of star formation,  where
the molecular content is more abundant, and spatially extended.
Daddi et al (2008) have optically selected
BzK galaxies, and detected in them much more CO emission than expected.
They are massive galaxies, with CO sizes up to 10kpc, with FIR luminosities
of 10$^{12}$ L$_\odot$. They have a normal SFE, with a high gas
content, M(H$_2$) $\sim$ 2  10$^{10}$ M$_\odot$, so that
the gas consumption time-scale is $\sim$ 2 Gyr, like the Milky Way.
 Their spatial extent explains why their CO-line excitation is low,
compared to nuclear starbursts.
In BzK-21000, for instance, at z=1.52, only weak CO(3-2) has been detected
(Dannerbauer et al  2009), and the CO-SLED is peaking at $J=3$ (as in the MW).
This leads to the choice of the standard CO-to-H$_2$ conversion factor,
 4.5 times higher than for ULIRGS. 
Due to selection effects, it is possible that SMG  studies
have missed a much larger population of gas-rich galaxies at high 
redshift.

\begin{figure}[b]
% \vspace*{-2.0 cm}
\begin{center}
 \includegraphics[width=11.2cm]{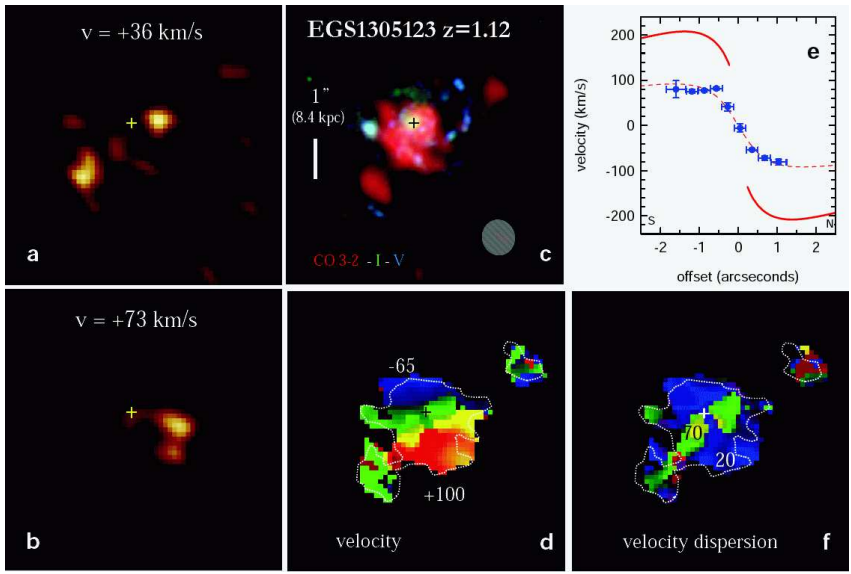} 
% \vspace*{-1.0 cm}
 \caption{
PdB CO(3-2) map at 0.6"x0.7" resolution, of the z=1.12 AEGIS galaxy EGS130512.
{\bf  a \& b:} two 9km/s channel maps at +36 and +72 km/s,
showing the massive molecular clumps, of typical gas masses $\sim$5x10$^9$ \msol,
intrinsic radii $<$1-2 kpc, gas surface densities 300-700 \msol.pc$^{-2}$ and velocity dispersions $\sim$19 km/s.
{\bf c:} CO integrated line emission (red), I-band (green) and V-band (blue) HST ACS images of the source.
{\bf d \& f:} peak velocity and velocity dispersion maps of the CO emission.
{\bf e:} peak CO velocity
along the major axis (PA=16$^\circ$). The best fitting exponential disk model with
radial scale length Rd=0.77" and dynamical mass of 2x10$^{11}$ \msol, for an adopted inclination
of 27$^\circ$ is shown
as a dotted red curve. The thick red curve is the deprojected rotation curve.
}
   \label{fig3}
\end{center}
\end{figure}

A recent survey of massive star forming galaxies, selected optically
to have comparable ranges of mass and SFR in the  
AEGIS and BX/BM surveys at z $\sim$ 1.2 and 2.3 was
reported by Tacconi et al (2010). About 20 galaxies 
were observed at IRAM-PdBI, with a
high detection rate $>$75\%, in these "normal" massive (M$_* \sim$  a few 10$^{11}$ M$_\odot$) Star
Forming Galaxies (SFG). The gas fraction appears to increase with redshift,
being in average 34\% and 44\% at z=1.2 and 2.3 respectively.
A typical map is shown in Figure \ref{fig3}, revealing a rotating disk,
perturbed by massive clumps.  The high gas content, with respect to local spirals,
implies more gravitational instabilities favoring
the formation of clumps.
 The survey prolongs the correlation between SFR versus stellar mass and redshift.
If the star formation rate was higher in the past, it is partly due
to the higher gas fraction in galaxies,
and also a little bit from a higher SF efficiency with z.
Given the short gas consumption time-scale in these objects,
galaxies must continuously accrete mass to explain the average gas fractions
as a function of time.

It is interesting to note the parallel mass assembly in galaxies 
and black holes, across time. Many of the objects detected in CO at high z
are AGN. 
Is there an influence of AGN feedback on CO emission?
A good example is the lensed quasar APM08279+5255 at z=3.9 
 (amplification factor $\sim$ 50).
This object is one of the brightest in the sky, and has been observed
with mm and cm telescopes.  Lines from
CO(1-0) to CO(11-10) have been detected.
Recent 0.3" resolution CO(1-0) mapping with VLA (Riechers et al 2009b)
reveals that the emission is compact, and not 
extended, as previously believed.  CO emission is 
co-spatial with optical/NIR,  in a circumnuclear disk of 500pc around
the black hole. The gas mass is 1.3 10$^{11}$ M$_\odot$, while the black hole mass is
M$_{BH}$ = 2.3 10$^{10}$M$_\odot$.
There is no hint of the influence of the AGN feedback. 
In this object, as many others at high z, the bulge mass
derived from the CO line-width, is 10 times less massive than the value
expected from the M$_{BH}$-$\sigma$ relation.

\begin{figure}[b]
% \vspace*{-2.0 cm}
\begin{center}
 \includegraphics[width=8.2cm]{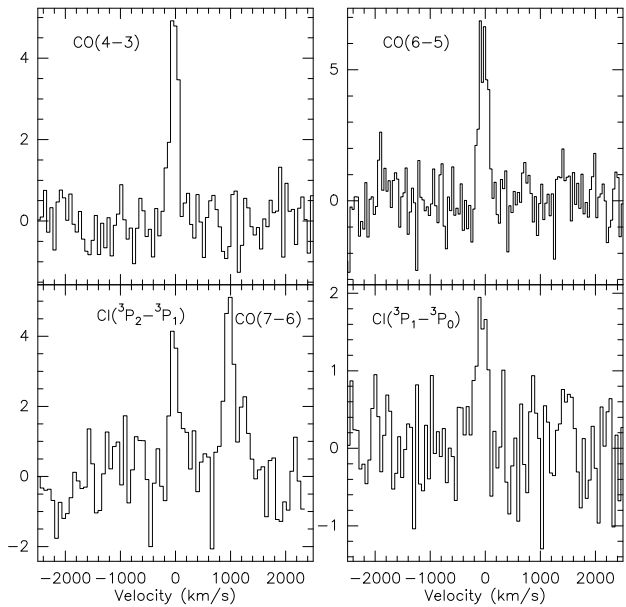} 
% \vspace*{-1.0 cm}
 \caption{
The three CO rotational lines detected, and the two \ci\, lines,
towards the brigtest SMG in the North, MM18423+5938, by Lestrade et al (2010).
The vertical scale is T$_{mb}$ in mK.
}
   \label{fig4}
\end{center}
\end{figure}

More statistically, Wang et al (2010) have displayed
the quasars studied at z $\sim$ 6, and also between
$1.4 < z < 5$, on the M$_{BH}$-$\sigma$ diagram,
revealing an order of magnitude
higher M$_{BH}$ than expected. However, some caveats must be taken
into account,  such as the unknown inclination, or other
selection biases.  For instance, it is easier to detect
the CO line in low-inclination objects, with a narrower
line-width.
ALMA will resolve the morphology of the CO emission, 
and determine actual inclinations.

An important progress has been made recently, thanks
to the enhanced bandwidth of millimeter receivers. It is
now possible to search for redshift with only the CO lines.
For objects with  z$>$2.2, there is always a CO line in the 
3mm atmospheric band of 80-116Ghz, since the spacing between lines decreases
as (1+z)$^{-1}$. When two lines are detected, it is possible
to deduce z, since the  frequency
difference between 2 lines depend on $J$.
Weiss et al (2009) have used this redshift machine to determine the redshift of
SMMJ14009+0252.
With the same techniques, Negrello et al (2010) and Lupu et al (2010) have 
determined the redshift of lensed SMG discovered behind
galaxy clusters, in the Herschel H-ATLAS survey.  At 500$\mu$m wavelength,
most sources with flux above 100mJy are lensed SMG, or nearby
spirals. In the Science Demonstration Phase,
11 sources were detected in 14 square degrees, and
6 were nearby galaxies. The remaining 5 were indeed lensed SMG,
and 4 of them were detected in CO.
Observations in the NIR domain give the nature of the lenses.
Extrapolating over the 550 square degrees of the survey,
 more than 100 lensed SMG will be discovered.

Using also the redshift machine at the
IRAM-30m, Lestrade et al (2010) have recently discovered the
brightest SMGs in the Northern hemisphere, quite serendipitously.
A source of 30mJy at 1.2mm with MAMBO was found, while searching for a debris-disk,
but without any local molecular cloud.
The redshift search with EMIR at IRAM-30m revealed the CO(6-5) and (4-3) lines,
together with the  2 CI lines, and CO(7-6) in
MM18423+5938, which is likely a 
lensed ULIRG galaxy, at z=3.93  (see Figure \ref{fig4}).
Another lensed star forming galaxy (magnification factor of 32) was discovered at 
z=2.33 by Swinbank et al (2010).
The size dilation due to the lens allows to 
resolve the star forming regions at z=2.33.
They are 100 times more luminous at a given size than
Giant Molecular Clouds in the Milky Way.

\section{Perspectives with ALMA}

The early science of ALMA will begin in 2011. In the near future,
ALMA will bring a breakthrough in our knowledge of the molecular
content of galaxies. First in continuum, deep fields will
give the source count N(S) to much lower fluxes, and
follow-up in the lines will precise the star formation history
of the universe. CO
lines will be intensively observed at high redshift, and
the gas fraction determined for "normal" systems.
This will allow to derive the efficiency of star formation
as a function of redshift, the gas kinematics, dynamical masses,
and clues to the various processes of galaxy formation.

%\begin{discussion}
%
%\discuss{J. QUESTION}{Are there ways to assess ..? 
%What are the signatures of such replenishment of gas?}
%
%\discuss{F. Combes}{It is difficult to find direct evidence, 
%}
%\end{discussion}

\end{document}